\pdfoutput=1
 \documentclass[aps,prl,twocolumn,showpacs,superscriptaddress,preprintnumbers]{revtex4-1}  
\usepackage[normalem]{ulem}
\usepackage{amsmath}
\usepackage{amssymb}
\usepackage{epsfig}
\usepackage{graphicx}
\usepackage{hyperref}
\usepackage{tabu}
\usepackage{boldline}
\usepackage{xspace}
\usepackage{slashed}
\usepackage{multirow}
\usepackage{diagbox}

\usepackage{color}
\usepackage[normal]{subfigure}
\usepackage[table]{xcolor}
\usepackage{enumitem}
\usepackage[utf8]{inputenc}
\usepackage{colortbl}
\definecolor{nicered}{rgb}{0.7,0.1,0.1}
\definecolor{nicegreen}{rgb}{0.1,0.5,0.1}
\definecolor{red}{rgb}{1.0, 0, 0}

\usepackage{lineno}
\definecolor{LightCyan}{rgb}{0.88,1,1}
\definecolor{piggypink}{rgb}{0.99, 0.87, 0.9}
\definecolor{applegreen}{rgb}{0.55, 0.71, 0.0}
\definecolor{darkpastelgreen}{rgb}{0.01, 0.75, 0.24}
\definecolor{green-yellow}{rgb}{0.68, 1.0, 0.18}

\newcommand{\beq}{\begin{equation}}
\newcommand{\eeq}{\end{equation}}
\newcommand{\beqa}{\begin{eqnarray}}
\newcommand{\eeqa}{\end{eqnarray}}

\newcommand{\unit}[1]{\ensuremath{\mathrm{\,#1}}\xspace}
\newcommand{\units}[1]{\unit{#1}}
\graphicspath{{figs/}}

\begin{document}
% ----------------- preprint numbers ------------------
% \begin{frontmatter}

% ------------- Title and authors ---------------------

\title{SENSEI: Direct-Detection Constraints on sub-GeV Dark Matter from \\ a Shallow Underground Run Using a Prototype Skipper-CCD}

\author{The SENSEI Collaboration: Orr Abramoff}
\affiliation{\normalsize\it 
Raymond and Beverly Sackler School of Physics and Astronomy, \\
 Tel-Aviv University, Tel-Aviv 69978, Israel}
 
\author{Liron Barak}
\affiliation{\normalsize\it 
Raymond and Beverly Sackler School of Physics and Astronomy, \\
 Tel-Aviv University, Tel-Aviv 69978, Israel}
 
 \author{Itay M. Bloch}
 \affiliation{\normalsize\it 
Raymond and Beverly Sackler School of Physics and Astronomy, \\
 Tel-Aviv University, Tel-Aviv 69978, Israel}
 
\author{Luke Chaplinsky}
\affiliation{\normalsize\it 
C.N.~Yang Institute for Theoretical Physics, Stony Brook University, Stony Brook, NY 11794}
\affiliation{\normalsize\it 
Department of Physics and Astronomy, Stony Brook University, Stony Brook University, Stony Brook, NY 11794} 
 
\author{Michael Crisler}
\affiliation{\normalsize\it 
Fermi National Accelerator Laboratory, PO Box 500, Batavia IL, 60510}

\author{Dawa}
\affiliation{\normalsize\it 
C.N.~Yang Institute for Theoretical Physics, Stony Brook University, Stony Brook, NY 11794}
\affiliation{\normalsize\it 
Department of Physics and Astronomy, Stony Brook University, Stony Brook University, Stony Brook, NY 11794} 

\author{Alex Drlica-Wagner}
\affiliation{\normalsize\it 
Fermi National Accelerator Laboratory, PO Box 500, Batavia IL, 60510}

 \author{Rouven Essig}
\affiliation{\normalsize\it 
C.N.~Yang Institute for Theoretical Physics, Stony Brook University, Stony Brook, NY 11794}

 \author{Juan Estrada}
\affiliation{\normalsize\it 
Fermi National Accelerator Laboratory, PO Box 500, Batavia IL, 60510}

\author{Erez Etzion}
\affiliation{\normalsize\it 
Raymond and Beverly Sackler School of Physics and Astronomy, \\
 Tel-Aviv University, Tel-Aviv 69978, Israel}

\author{Guillermo Fernandez}
\affiliation{\normalsize\it 
Fermi National Accelerator Laboratory, PO Box 500, Batavia IL, 60510}

\author{Daniel Gift}
\affiliation{\normalsize\it 
C.N.~Yang Institute for Theoretical Physics, Stony Brook University, Stony Brook, NY 11794}
\affiliation{\normalsize\it 
Department of Physics and Astronomy, Stony Brook University, Stony Brook University, Stony Brook, NY 11794} 

 \author{Joseph Taenzer}
\affiliation{\normalsize\it 
Raymond and Beverly Sackler School of Physics and Astronomy, \\
 Tel-Aviv University, Tel-Aviv 69978, Israel}
 
 \author{Javier Tiffenberg}
\affiliation{\normalsize\it 
Fermi National Accelerator Laboratory, PO Box 500, Batavia IL, 60510}

 \author{Miguel Sofo Haro}
\affiliation{\normalsize\it 
Fermi National Accelerator Laboratory, PO Box 500, Batavia IL, 60510}
\affiliation{Centro At\'omico Bariloche, CNEA/CONICET/IB, Bariloche, Argentina}

 \author{Tomer Volansky}
\affiliation{\normalsize\it 
Raymond and Beverly Sackler School of Physics and Astronomy, \\
 Tel-Aviv University, Tel-Aviv 69978, Israel}
 \affiliation{\normalsize\it
 School of Natural Sciences, The Institute for Advanced Study, Princeton, NJ 08540, USA}

 \author{Tien-Tien Yu}
\affiliation{\normalsize\it 
Department of Physics and Institute of Theoretical Science, University of Oregon, Eugene, Oregon 97403}

\preprint{YITP-2019-01, FERMILAB-PUB-19-039-AE}
% ------------------------------------------------------
\begin{abstract}
We present new direct-detection constraints on eV-to-GeV dark matter interacting with electrons using a prototype detector of the Sub-Electron-Noise Skipper-CCD Experimental Instrument. The results are based on data taken in the MINOS cavern at the Fermi National Accelerator Laboratory.  We focus on data obtained with two distinct readout strategies.  For the first strategy, we read out the Skipper-CCD continuously, accumulating an exposure of 0.177~gram-days.  While we observe no events containing three or more electrons, we find a large one- and two-electron background event rate, which we attribute to spurious events induced by the amplifier in the Skipper-CCD readout stage. For the second strategy, we take five sets of data in which we switch off all amplifiers while exposing the Skipper-CCD for 120k seconds, and then read out the data through the best prototype amplifier.  We find a one-electron event rate of $(3.51\pm 0.10)\times 10^{-3}$~events/pixel/day,  which is almost two orders of magnitude lower than the one-electron event rate observed in the  continuous-readout data, and a two-electron event rate of $(3.18^{+0.86}_{-0.55})\times 10^{-5}$~events/pixel/day.  We again observe no events containing three or more electrons, for an exposure of 0.069~gram-days. We use these data to derive world-leading constraints on dark matter-electron scattering for masses between 500~keV to 5~MeV, and on dark-photon dark matter being absorbed by electrons for a range of masses below 12.4~eV. 
\end{abstract}

\maketitle

%%%%%%%%%%%%%%%%%%%%%%%%%%%%%%%%%%%%%%%%%%%%%%%%%%%%%%%%%%%%%%%%%%%%%%%%%%%%
%%%%%%%%%%%%%%%%%%%%%%%%%%%%%%%%%%%%%%%%%%%%%%%%%%%%%%%%%%%%%%%%%%%%%%%%%%%%
%%%%%%%%%%%%%%%%%%%%%%%%%%%%%%%%%%%%%%%%%%%%%%%%%%%%%%%%%%%%%%%%%%%%%%%%%%%%%%%%%%%%%%%%%%%%%%%%%%%%%%%%%%%%%%%%%%%%%%%%%%%%%%%%%%%%%%%%%%%%%%%%%%%%%%%%
\textbf{INTRODUCTION.}  
Direct-detection experiments play an important
role in our quest to understand the nature of dark matter (DM). 
While Weakly Interacting Massive Particles (WIMPs), with masses in the 10~GeV-1~TeV range, have been the main target of these experiments, the existence of well-motivated DM candidates with eV-to-GeV masses has prompted theoretical and experimental efforts to probe also this lower mass range~\cite{Battaglieri:2017aum}.  This is possible, for example, by searching for DM interactions with electrons in various target materials~\cite{Essig:2011nj}. 

The Sub-Electron-Noise Skipper-CCD Experimental Instrument (SENSEI) is the first dedicated direct-detection experiment to search for DM in the eV-to-GeV range.  SENSEI uses silicon Skipper-Charge-Coupled-Devices (Skipper-CCDs) consisting of $\mathcal{O}$(million) pixels.  
An electron in the valence band of the silicon can be promoted to the conduction band after it scatters off a 
DM particle with mass above $\sim$500~keV, or after it absorbs a DM particle with mass above the silicon band gap of $\sim$1.2~eV~\cite{Essig:2011nj,Essig:2015cda,Lee:2015qva,Graham:2012su,An:2014twa,Bloch:2016sjj,Hochberg:2016sqx}. The excited electron subsequently relaxes to the bottom of the conduction band, creating an 
additional electron-hole pair for each 3.8~eV of excitation energy above the band gap~\cite{Vavilov1962}. 
The resulting electron-hole pairs~\footnote{Below, we refer to ``electron-hole pairs'' simply as ``electrons''.} are moved pixel-by-pixel to one of the Skipper-CCD corners containing the ultralow-noise readout stages that measure precisely their number~\cite{Tiffenberg:2017aac}.
Throughout this paper, we refer to each contiguous grouping of pixels containing one or more electrons as an `event.'

We present here results from a prototype Skipper-CCD placed in the MINOS cavern, located about 104~m~\cite{Adamson:2007gu} underground at the Fermi National Accelerator Laboratory (FNAL). 
We demonstrate how the Skipper-CCD can be operated in different running modes, allowing us to identify a previously unknown source of background, which arises from soft photons emitted by an amplifier operated at the readout stage.  We present the resulting DM constraints, and discuss the impact of different running modes on the observed instrumental background of one- and two-electron events.  
Results from a surface run with this prototype were presented in~\cite{Crisler:2018gci}. 

%%%%%%%%%%%%%%%%%%%%%%%%%%%%%%%%%%%%%%%%%%%%%%%%%%%%%%%%%%%%%%%%%%%%%%%%%%%%
%%%%%%%%%%%%%%%%%%%%%%%%%%%%%%%%%%%%%%%%%%%%%%%%%%%%%%%%%%%%%%%%%%%%%%%%%%%%
%%%%%%%%%%%%%%%%%%%%%%%%%%%%%%%%%%%%%%%%%%%%%%%%%%%%%%%%%%%%%%%%%%%%%%%%%%%%
%%%%%%%%%%%%%%%%%%%%%%%%%%%%%%%%%%%%%%%%%%%%%%%%%%%%%%%%%%%%%%%%%%%%%%%%%%%%
\textbf{THE SENSEI PROTOTYPE DETECTOR.}
The SENSEI prototype detector (``protoSENSEI'') consists of a single Skipper-CCD placed in a light-tight copper housing, with an active area of 1.086~cm $\times$ 1.872~cm and a total active mass (before masking) of 0.0947~gram. 
This prototype sensor has four amplifiers as part of the readout stage located in the four corners of the Skipper-CCD, each with a distinct design and noise performance.  
The four amplifiers allow the Skipper-CCD to be divided into four equal-sized quadrants, each consisting of 624 rows of 362 pixels.  Each pixel has an area of $15~\mu\textrm{m} \times 15~\mu\textrm{m}$, a thickness of 200~$\mu$m, and a mass of $1.0476\times 10^{-7}$~gram.
While each amplifier usually reads one quadrant, it is also possible to have two quadrants read out through one amplifier. 
One amplifier design has high noise from charge-misclassification problems, and we discarded all data from it. 
The electronics consists of a modified Monsoon system as described in~\cite{Tiffenberg:2017aac}. 
All data presented below were obtained by measuring each pixel 800 times.

\begin{figure}[t!]
\begin{center}
\includegraphics[width=0.48\textwidth]{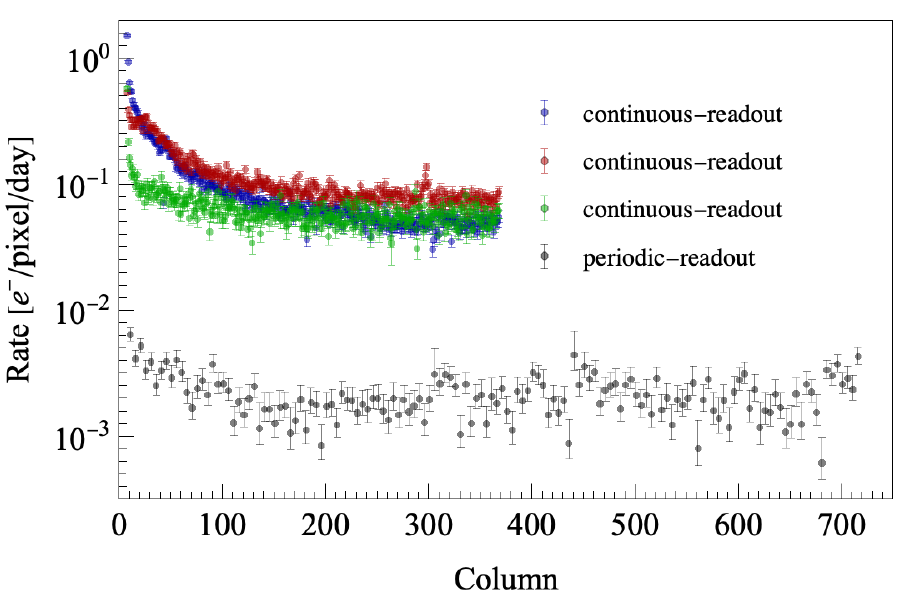}
\caption{The average measured one-electron rate 
as a function of the Skipper-CCD column number for the \texttt{continuous-readout} data (top three colored data sets, shown for the three working amplifiers) and for the \texttt{periodic-readout} data, in which two quadrants are read through the best amplifier (bottom black data set; the latter was averaged and binned for five columns).  Infrared photons produced by the amplifier (located near column~0) lead to an increasing background event rate in columns close to the amplifier. 
For the \texttt{periodic-readout}, another amplifier is also located near column 724, but was turned off at all times. 
} 
\label{fig:DCrate}
\end{center}
\end{figure}

%%%%%%%%%%%%%%%%%%%%%%%%%%%%%%%%%%%%%%%%%%%%%%%%%%%%%%%%%%%%%%%%%%%%%%%%%%%%
%%%%%%%%%%%%%%%%%%%%%%%%%%%%%%%%%%%%%%%%%%%%%%%%%%%%%%%%%%%%%%%%%%%%%%%%%%%%
%%%%%%%%%%%%%%%%%%%%%%%%%%%%%%%%%%%%%%%%%%%%%%%%%%%%%%%%%%%%%%%%%%%%%%%%%%%%
%%%%%%%%%%%%%%%%%%%%%%%%%%%%%%%%%%%%%%%%%%%%%%%%%%%%%%%%%%%%%%%%%%%%%%%%%%%%
\textbf{DATA COLLECTION STRATEGIES.}
We took various sets of data with different exposure times and readout modes to understand detector backgrounds and constrain DM interactions. 
The largest dataset was collected by reading the Skipper-CCD ``continuously'' and in parallel with four amplifiers, with each amplifier reading a single quadrant independently. The exposure time of each pixel is given by the Skipper-CCD readout time, which is about 4.4k seconds. 
We noticed that all quadrants have an increasing density of one- and two-electron events closer to the long edge of the Skipper-CCD, where the amplifier is located, suggesting that the amplifier is producing excess events.  
When reading, the voltages on the amplifier are adjusted rapidly while sampling the charge packet in each pixel. These voltage variations increase the base current of the amplifier, which increases the probability of producing infrared photons~\cite{Toriumi:1486823}.
These infrared photons can reach the active Skipper-CCD region and contribute to spurious events with decreasing profile towards the center of the detector, see Fig.~\ref{fig:DCrate}.
Since all rows are exposed for the same time to the amplifier when reading continuously, we first sum across all images in our dataset (discussed further below) the total number of one-electron events 
in a given column, and divide by the total number of pixels.  The figure shows the resulting mean number of one-electron events per pixel per day as a function of the column index for each of the three working amplifiers. 
We distinguish these ``amplifier-induced events'' from the ``dark current'', namely few-electron events that are distributed evenly across the Skipper-CCD region and are due to thermal fluctuations that occasionally promote valence-band electrons to the conduction band.  

The observed rate of one-electron events depends on the distance from the amplifier, but is of order  $\sim$0.1~events/pixel/day.  We can still use the \texttt{continuous-readout} data to perform a background-free search for events with three to 100 electrons.  

In order to reduce the excess of events from the amplifier, we took data by exposing the Skipper-CCD for some time before reading it out periodically.  During the exposure, the amplifiers were on but not actively moving or measuring charge packets, and we find that exposures of 30k--60k seconds reduce the one-electron rate by about an order of magnitude compared to the \texttt{continuous-readout} mode.  We then took data for which the amplifiers were turned off during the exposure and only turned on to read out the data. We find another order-of-magnitude reduction in the one-electron rate compared to the periodic-readout, amplifier-on data, suggesting that 
the amplifier is constantly emitting infrared photons, with the excess largest during readout.  This is expected, since the base current is larger when the amplifier is actively reading.  Finally, we took data with the amplifier off during the exposure for which we read out two quadrants through the amplifier with the best single-sample noise properties (yielding images with twice as many columns), see Fig.~\ref{fig:DCrate}. While this did not reduce further the one-electron rate, we will present results below using five sets of data obtained in this readout mode, which we call \texttt{periodic-readout}, each with an exposure of 120k seconds.  

In summary, we use below a \texttt{continuous-readout} dataset to constrain the event rate of three and more electrons, and a \texttt{periodic-readout} dataset to constrain also the one- and two-electron event rates.  Since the detector is located at a shallow site, the image occupancy of the periodic-readout dataset, from multiple high-energy events, is high.

%%%%%%%%%%%%%%%%%%%%%%%%%%%%%%%%%%%%%%%%%%%%%%%%%%%%%%%%%%%%%%%%%%%%%%%%%%%%
%%%%%%%%%%%%%%%%%%%%%%%%%%%%%%%%%%%%%%%%%%%%%%%%%%%%%%%%%%%%%%%%%%%%%%%%%%%%
%%%%%%%%%%%%%%%%%%%%%%%%%%%%%%%%%%%%%%%%%%%%%%%%%%%%%%%%%%%%%%%%%%%%%%%%%%%%
%%%%%%%%%%%%%%%%%%%%%%%%%%%%%%%%%%%%%%%%%%%%%%%%%%%%%%%%%%%%%%%%%%%%%%%%%%%%
\textbf{DATA QUALITY CUTS.}
For both datasets, we use the same event-selection criteria as described in~\cite{Crisler:2018gci}, together with additional quality cuts developed from our improved understanding of low-energy backgrounds and detector effects:
\begin{itemize}[leftmargin=*]\addtolength{\itemsep}{-0.6\baselineskip}
\item {\bf Single-pixel events \& neighbour mask.} To simplify our analysis and reject multi-pixel events produced by random coincidence of one-electron pixels, we require the DM signal to be contained in a single pixel and only select pixels whose neighboring pixels are empty~\cite{Crisler:2018gci}. 
%
%%%
\begin{figure}[t!]
\begin{center}
\includegraphics[width=0.49\textwidth]{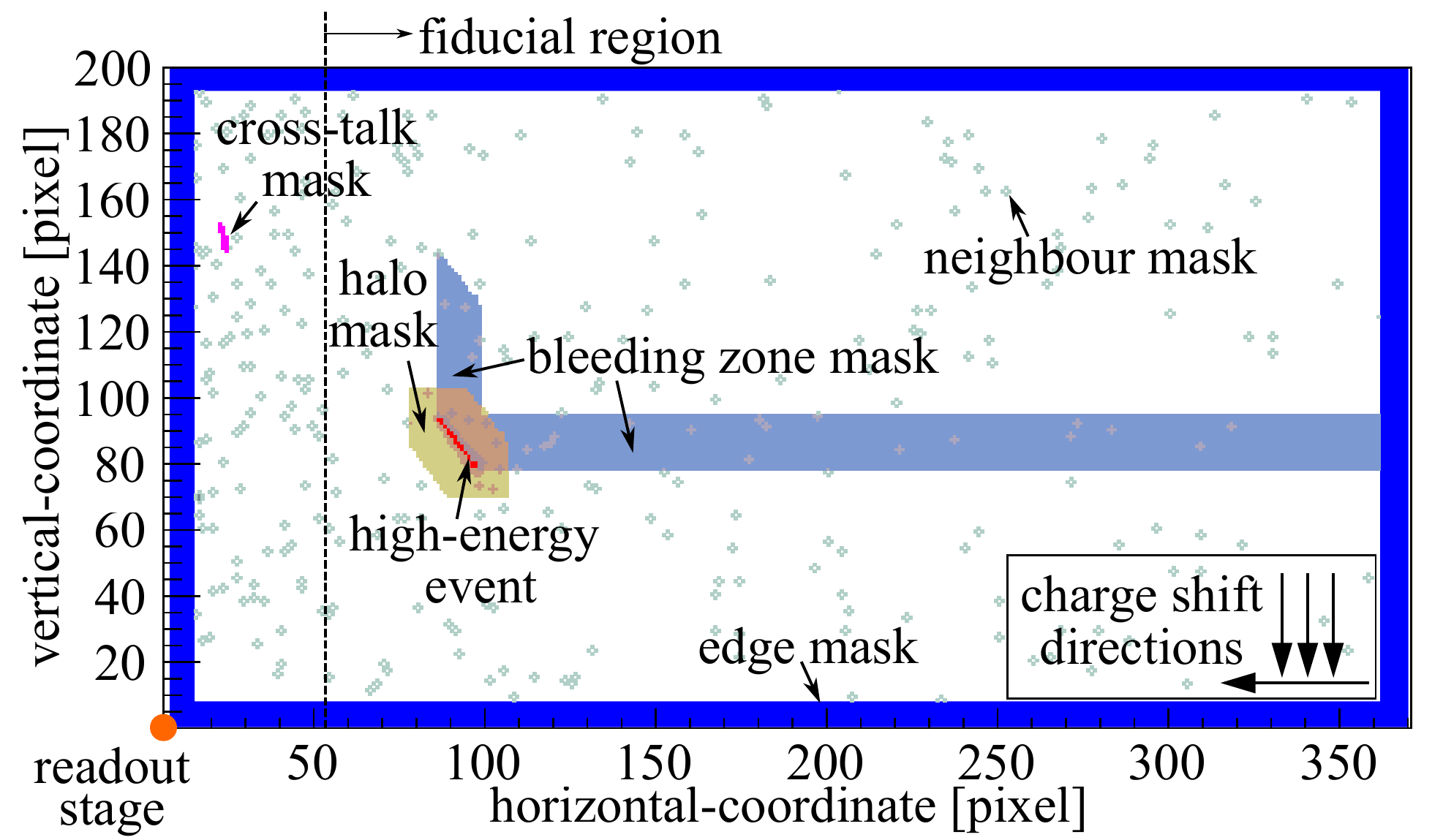}
\caption{Example of an image consisting of 200~rows from the \texttt{continuous-readout} data showing events and the mask after applying data-quality cuts. (The bad-column cut is not shown for presentation purposes only.)
}
\label{fig:image}
\end{center}
\end{figure}
%%%
\item {\bf Electronic noise.}  We veto images in which the readout noise is 30\% larger than the expected readout noise as inferred from an over-scan region in which virtual (non-existent) pixels are read.  
\item {\bf Edge mask.}  We remove 8 pixels around the edge of each image to avoid any edge effects due to the non-uniformity in the electric field~\cite{Plazas:2014aha}. 
\item {\bf Bleeding zone mask.} 
Due to charge-transfer inefficiency, we mask 50 pixels upstream in the vertical and horizontal direction of any pixel containing more than 100 electrons in the \texttt{periodic-readout} data. 
However, we observe that the 
the charge-transfer inefficiency is slightly larger for charge-shifts on the serial register (horizontal).  To avoid any chance of spurious events with three or more electrons in the \texttt{continuous-readout} data (consisting of low-occupancy images), we reject all pixels in the horizontal direction of pixels with more than 100 electrons (in addition to the 50 pixels in the vertical direction). 
\item {\bf Halo mask.} The pixels around events with many electrons show an increased rate of low-energy events. We plan to study these in future work, but suspect that they are produced by infrared photons created by bremsstrahlung or recombination of ionized electrons. Here, we reject events located less than eight pixels away from any pixel containing more than 100 electrons. 
\item {\bf Cross-talk mask.} 
When reading the four quadrants simultaneously, high-energy signals recorded in one of the four quadrants can produce a fake signal in one or more of the other three due to electronic cross talk~\cite{Bernstein:2017gsy}. We reject pixels from all four quadrants that were read at the same time as a pixel containing more than 500~electrons.
\item {\bf Bad columns.} Some pixels could contain defects or impurities that cause charge leakage, especially if a high-energy event occurs near such pixels.  These appear as columns that contain an excess of non-empty pixels.  To identify affected columns while keeping the analysis blind,  we analyzed multiple sets of commissioning  data exposing the Skipper-CCD for 120k seconds each.  Moreover, we analyzed the 50 masked pixels immediately above the high-energy events in the data used to derive our limits. Altogether we conservatively discard 60 ``bad columns'' whose noise was more than 2.5 standard deviations above the median noise across all columns averaged over all images. 
\end{itemize}%

\begin{table}[t]
\begin{center}
\begin{tabular}{|l||c|c|c||c|c|c|}
\hline
\multirow{2}{*}{\diagbox{Cuts}{ $N_{e}$}} & \multicolumn{3}{c||}{\texttt{periodic}} & \multicolumn{3}{c|}{\texttt{continuous}} \\ \cline{2-7}
                            & 1      & 2    &  3        & 3       & 4       & 5         \\ \hline \hline
1.~DM in single pixel       & 1      & 0.62 & 0.48      & 0.48    & 0.41    & 0.36      \\ \hline
2.~Nearest Neighbour        &\multicolumn{3}{c||}{0.92} & \multicolumn{3}{c|}{0.96}     \\ 
3.~Electronic Noise         &\multicolumn{3}{c||}{1   } & \multicolumn{3}{c|}{$\sim$1}  \\ 
4.~Edge                     &\multicolumn{3}{c||}{0.92} & \multicolumn{3}{c|}{0.88}     \\ 
5.~Bleeding                 &\multicolumn{3}{c||}{0.71} & \multicolumn{3}{c|}{0.98}     \\ 
6.~Halo                     &\multicolumn{3}{c||}{0.80} & \multicolumn{3}{c|}{0.99}     \\ 
7.~Cross-talk               &\multicolumn{3}{c||}{0.99} & \multicolumn{3}{c|}{$\sim$1}  \\ 
8.~Bad columns              &\multicolumn{3}{c||}{0.80} & \multicolumn{3}{c|}{0.94}     \\ \hline
Total Efficiency                      &  0.38  & 0.24 & 0.18      & 0.37    & 0.31    & 0.28      \\ \hline
Eff.~Expo.~[g day]          &  0.069 & 0.043& 0.033     & 0.085   & 0.073   & 0.064     \\ \hline\hline
%Eff.~Expo. [g-day]          &  0.07  & 0.04 & 0.03      & 0.09    & 0.07    & 0.06     \\ \hline\hline
Number of events            & 2353   & 21   & 0         & 0       & 0       & 0         \\ \hline
\end{tabular}
\caption{Efficiencies for the data-selection cuts for the \texttt{periodic-readout} and \texttt{continuous-readout} datasets, for events with 1 to 5 electrons. The bottom two rows list the the efficiency-corrected exposures and the number of observed events after cuts, respectively. 
\vspace{-6mm}
} 
\label{tab:eff}
\end{center}
\end{table}%

\begin{figure*}[t]
\begin{center}
 \includegraphics[width=0.49\textwidth]{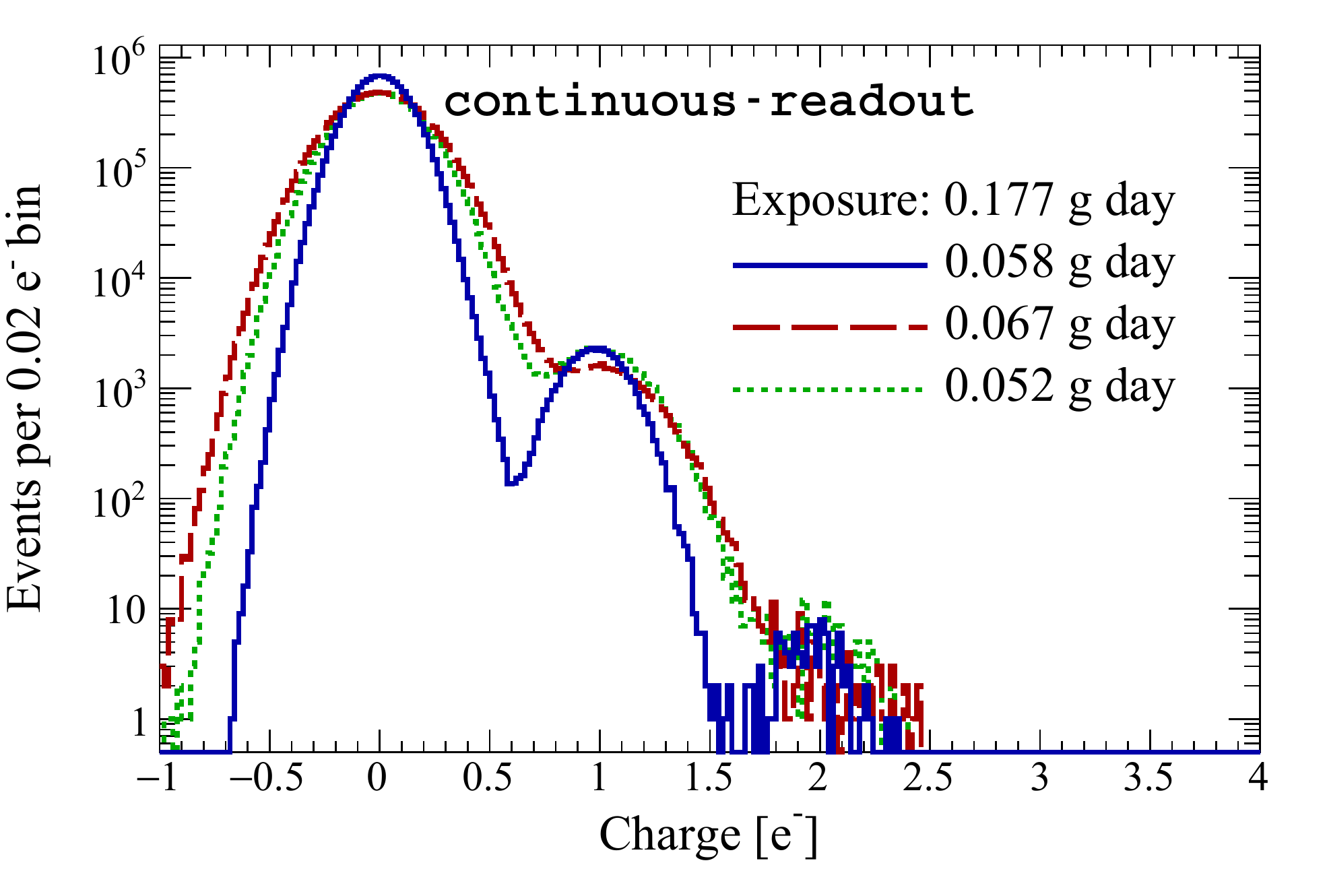}
 \includegraphics[width=0.49\textwidth]{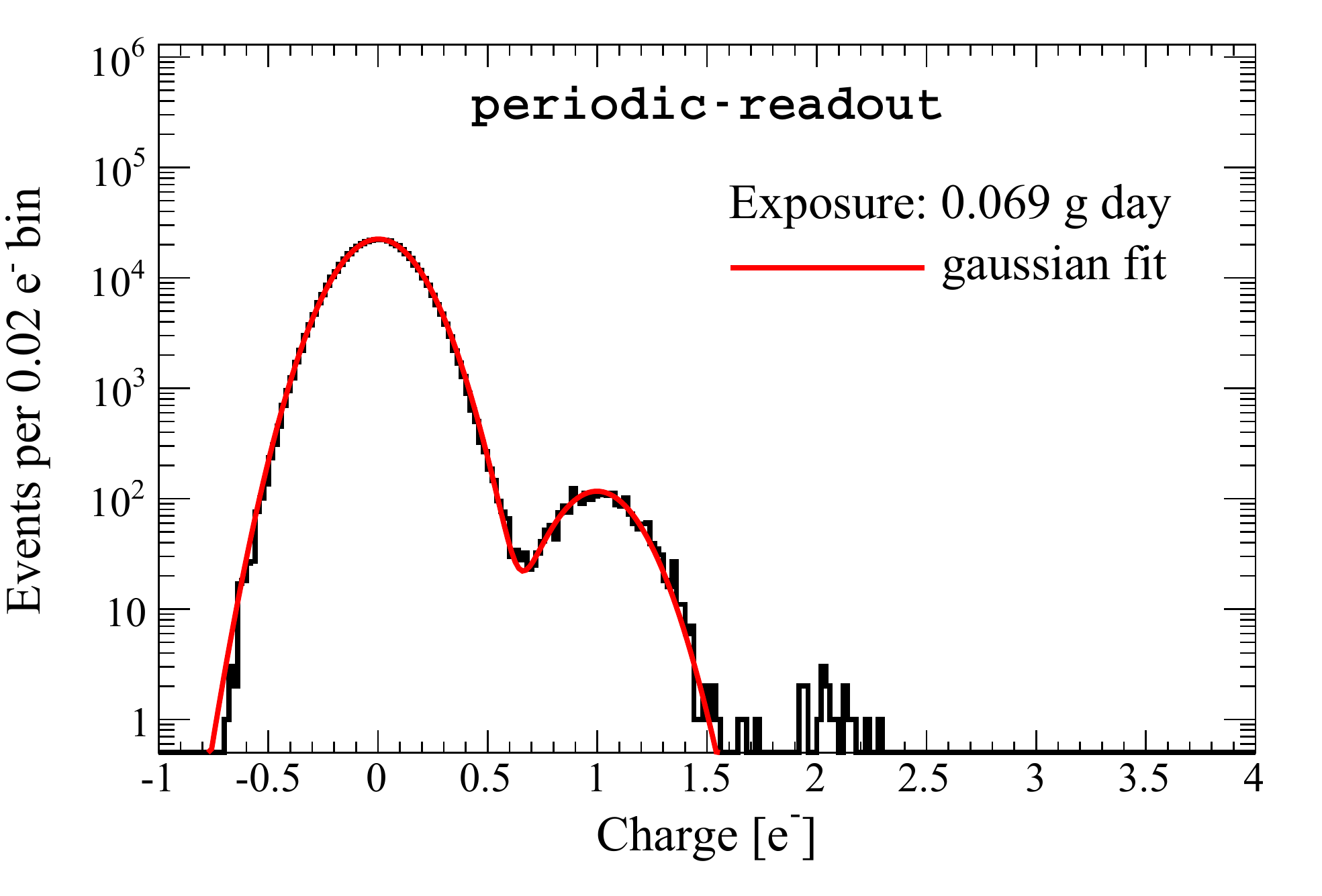}
\caption{ Spectra of the recorded events for the \texttt{continuous-readout} (\textbf{left}) and \texttt{periodic-readout} (\textbf{right}) data.
For the \texttt{continuous-readout} data, we show the spectra recorded by the three working amplifiers. 
The widths of the charge distributions depend on the amplifier design. 
The \texttt{periodic-readout} spectrum corresponds to the total number of events found in the eight double-quadrant images used to constrain the rate of events containing two and more electrons. 
There are no events with measured charge greater than 2.5~electrons in either data. Exposures include all efficiencies except for ``Cut~1'' from Table~\ref{tab:eff}.
}
\label{fig:spectra}
\end{center}
\end{figure*}

All cuts above were developed on commissioning data or in the masked region of the physics data, and then applied to the unmasked physics data regions. An example of a masked image (but without the bad-column cut) is shown in Fig.~\ref{fig:image}.  
A summary of selection efficiencies 
are listed in Table~\ref{tab:eff} for electron bins $1-5$. Bins $1-3$ provide the best constraint on DM-electron scattering, while we use bins $1-100$ to constrain DM absorption. 
The number of events with a given number of electrons is determined from a fit, since at only 800~samples per pixel in this prototype sensor there is some smearing from neighbouring electron bins.  
We now discuss the analysis of each dataset in turn.

%%%%%%%%%%%%%%%%%%%%%%%%%%%%%%%%%%%%%%%%%%%%%%%%%%%%%%%%%%%%%%%%%%%%%%%%%%%%
%%%%%%%%%%%%%%%%%%%%%%%%%%%%%%%%%%%%%%%%%%%%%%%%%%%%%%%%%%%%%%%%%%%%%%%%%%%%
%%%%%%%%%%%%%%%%%%%%%%%%%%%%%%%%%%%%%%%%%%%%%%%%%%%%%%%%%%%%%%%%%%%%%%%%%%%%
%%%%%%%%%%%%%%%%%%%%%%%%%%%%%%%%%%%%%%%%%%%%%%%%%%%%%%%%%%%%%%%%%%%%%%%%%%%%

\textbf{CONTINUOUS-READOUT DATA ANALYSIS.}
These data consist of images from the three working amplifier taken over 3.8~days, corresponding to 0.27~g-day, which we use to constrain the rate of events containing three to 100 electrons. 

Despite the excess events being produced by the amplifier, we can fiducialize the images by removing pixels that are too close to the amplifier and find the optimal constraint on the three-electron event rate. 
To do this, we must remove several columns close to the amplifier. We design the optimal column-cut (after masking) using Fig.~\ref{fig:DCrate}. 
We assume that the excess events produced by the amplifier follows a Poisson distribution, and predict the number of three-electron events that would remain in the entire dataset as a function of the column index.  
We find that the minimum column indices for the three amplifiers that maximize the total exposure time and predict not more than 0.5 three-electron events are 55, 10, and 53, respectively. 
After applying these column-cuts, we unblind and find the spectra shown in Fig.~\ref{fig:spectra} (left).  We find zero events with three (or more) electrons in the unblinded data. The final exposures after all data cuts (in g-day) for each quadrant are 0.058, 0.067, and 0.052, respectively, for a total of 0.177~g-day.

%%%%%%%%%%%%%%%%%%%%%%%%%%%%%%%%%%%%%%%%%%%%%%%%%%%%%%%%%%%%%%%%%%%%%%%%%%%%
%%%%%%%%%%%%%%%%%%%%%%%%%%%%%%%%%%%%%%%%%%%%%%%%%%%%%%%%%%%%%%%%%%%%%%%%%%%%
%%%%%%%%%%%%%%%%%%%%%%%%%%%%%%%%%%%%%%%%%%%%%%%%%%%%%%%%%%%%%%%%%%%%%%%%%%%%
%%%%%%%%%%%%%%%%%%%%%%%%%%%%%%%%%%%%%%%%%%%%%%%%%%%%%%%%%%%%%%%%%%%%%%%%%%%%
\textbf{PERIODIC-READOUT DATA ANALYSIS.}
We took five sets of 120k-second-exposure, double-quadrant-readout data.  After applying the data-quality cuts, each dataset is divided into three images of 200 rows each.  
To constrain the one-electron event rate, we apply additional data-selection criteria, which were determined from analyzing other  120k-seconds-exposure data.  First, we remove all five images that were read out last, since these have the longest exposure to the amplifier during readout.  We then calculate the rate of events containing five or fewer electrons inside the masked regions of the remaining ten images, which we found in commissioning data to be positively correlated with the one-electron event rate outside the masked regions.  We took the four images with the lowest rate in the masked region, and then measured their average one-electron event rate outside the masked regions, finding $(3.51 \pm 0.10)\times 10^{-3}$~events/pixel/day, with a 90\%~CL~upper limit of $3.68 \times 10^{-3}$~events/pixel/day.  

\begin{figure*}[t]
\begin{center}
 \includegraphics[width=0.32\textwidth]{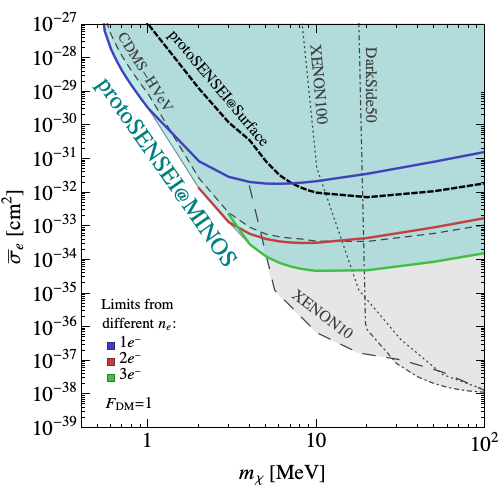}
 \includegraphics[width=0.32\textwidth]{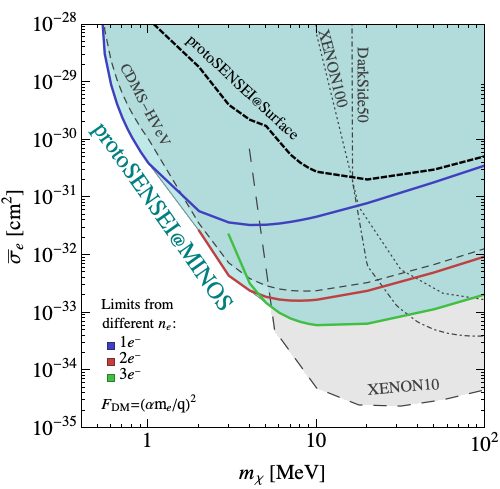}
\includegraphics[width=0.32\textwidth]{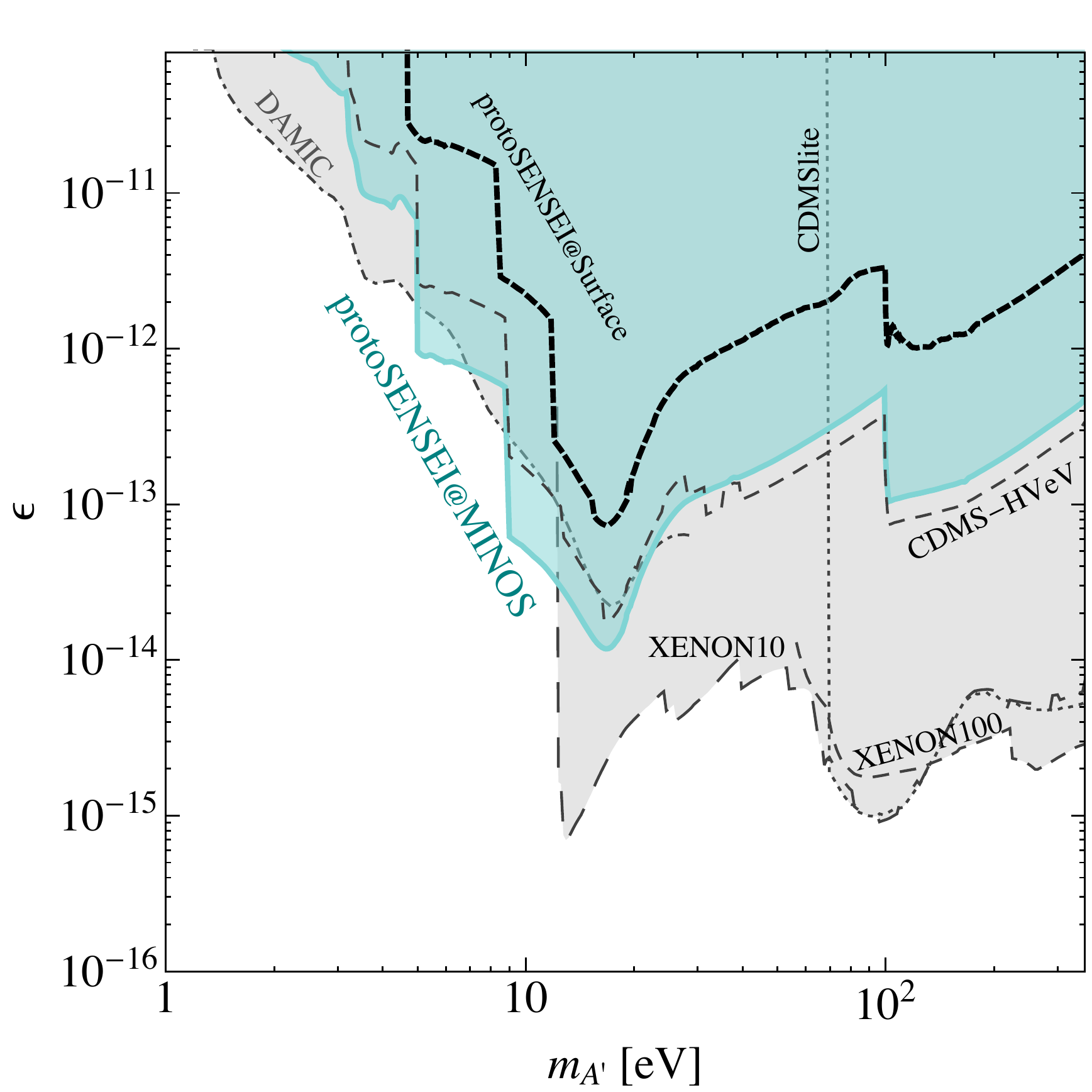}
\caption{
The 90\%~CL~constraints (cyan shaded regions) from a SENSEI prototype detector located underground near the MINOS cavern at FNAL.  We show constraints on the DM-electron scattering cross-section, $\overline{\sigma}_e$, 
as a function of DM mass, $m_\chi$, for two different DM form factors, $F_{\rm DM}(q)=1$ (\textbf{left}) and $F_{\rm DM}(q)=(\alpha m_e/q)^2$ (\textbf{middle}), and constraints on the kinetic-mixing parameter, $\epsilon$, versus the dark-photon mass, $m_{A'}$, for dark-photon-dark-matter absorption (\textbf{right}). 
The thick blue and red lines use the one- and two-electron-rate constraints from the \texttt{periodic-readout} data, respectively, while the green line combines the three-electron-rate constraints from the \texttt{continuous-readout} and \texttt{periodic-readout} data.  
Constraints are also shown from the SENSEI surface run~\cite{Crisler:2018gci}, XENON10 and XENON100~\cite{Essig:2017kqs}, DarkSide-50~\cite{Agnes:2018oej}, and CDMS-HVeV (without Fano-factor fluctuations)~\cite{Agnese:2018col} for the left and middle plots, and from the SENSEI surface run~\cite{Crisler:2018gci}, DAMIC~\cite{Aguilar-Arevalo:2016zop}, XENON10, XENON100, and CDMSlite~\cite{Bloch:2016sjj} for the absorption limits on the right plot. 
}
\label{fig:DMresults}
\end{center}
\end{figure*}

To constrain the two-electron event rate, 
we take the observed number of one-electron events in each of the ten images closest to the amplifier, and calculate the expected number of two-electron events in each of these images, assuming 
a Poisson distribution. 
We find that including the eight images with the lowest one-electron rate yields the lowest expected 90\% CL~limit on the two-electron event rate, and an expected $\sim$6.5 two-electron events. 
After unblinding these eight images, we find 21 events and a two-electron event rate of $(3.18^{+0.86}_{-0.55})\times 10^{-5}$~events/pixel/day.  This is more than expected, which we find is attributable to an insufficient masking of these high-occupancy images.  Nevertheless, we include all observed two-electron events to find a 90\%~CL~upper limit of $4.27\times 10^{-5}$~two-electron-events/pixel/day. The measured exposure (after all cuts) is 0.069 g-day.  
The observed spectrum of events from these eight images is shown in Fig.~\ref{fig:spectra} (right). We see no events with three to 100  electrons, and add this \texttt{periodic-readout} data to the \texttt{continuous-readout} data to constrain DM that produces three to 100 electrons, for a combined exposure of 0.246~g-day.  

%%%%%%%%%%%%%%%%%%%%%%%%%%%%%%%%%%%%%%%%%%%%%%%%%%%%%%%%%%%%%%%%%%%%%%%%%%%%
%%%%%%%%%%%%%%%%%%%%%%%%%%%%%%%%%%%%%%%%%%%%%%%%%%%%%%%%%%%%%%%%%%%%%%%%%%%%
%%%%%%%%%%%%%%%%%%%%%%%%%%%%%%%%%%%%%%%%%%%%%%%%%%%%%%%%%%%%%%%%%%%%%%%%%%%%
%%%%%%%%%%%%%%%%%%%%%%%%%%%%%%%%%%%%%%%%%%%%%%%%%%%%%%%%%%%%%%%%%%%%%%%%%%%%
\textbf{DARK MATTER RESULTS.}
In Fig.~\ref{fig:DMresults}, we show 90\%~CL~upper limits on the  DM-electron scattering cross section ~\cite{Essig:2011nj,Essig:2015cda} and dark-photon dark matter absorption~\cite{An:2014twa,Bloch:2016sjj,Hochberg:2016sqx}. 
We assume a local DM density of $\rho_{\rm DM}=0.3\units{GeV/cm^3}$~\cite{Bovy:2012tw}, a standard isothermal Maxwellian velocity distribution~\cite{Lewin:1995rx} with a DM escape velocity of 544~km/s, and a mean local velocity of 220~km/s.  
To be conservative, we do not include Fano-factor fluctuations. 

For DM-electron scattering, $m_\chi\lesssim 1~\textrm{MeV}$ ($1~\textrm{MeV}\lesssim m_\chi\lesssim 4~\textrm{MeV}$) is constrained most stringently by the observed one-electron (two-electron) event rate in the \texttt{periodic-readout} data, while the combined \texttt{continuous-readout} and \texttt{periodic-readout} data provides the best SENSEI constraint for $m_\chi>0.4~\textrm{MeV}$ from observing no three-electron events. These results provide the most stringent direct-detection constraints on DM-electron scattering for $500~\textrm{keV}\lesssim m_\chi \lesssim 5~\textrm{MeV}$. 
For DM absorption, SENSEI now provides the world-leading constraint for some range of masses below 12.4~eV. 
%\TY{We re-analyze the CDMS HV-eV data through our pipeline in order to compare absorption curves to protoSENSEI absorption.}

%%%%%%%%%%%%%%%%%%%%%%%%%%%%%%%%%%%%%%%%%%%%%%%%%%%%%%%%%%%%%%%%%%%%%%%%%%%%
%%%%%%%%%%%%%%%%%%%%%%%%%%%%%%%%%%%%%%%%%%%%%%%%%%%%%%%%%%%%%%%%%%%%%%%%%%%%
%%%%%%%%%%%%%%%%%%%%%%%%%%%%%%%%%%%%%%%%%%%%%%%%%%%%%%%%%%%%%%%%%%%%%%%%%%%%
%%%%%%%%%%%%%%%%%%%%%%%%%%%%%%%%%%%%%%%%%%%%%%%%%%%%%%%%%%%%%%%%%%%%%%%%%%%%
\textbf{OUTLOOK.} 
The SENSEI Collaboration is procuring $\sim 100$~grams of new Skipper-CCDs and custom-designing electronics for an experiment at SNOLAB. We expect these new sensors to have an improved noise performance and lower dark-count rate due to the use of higher-quality silicon. 
We are implementing mitigation strategies for amplifier-induced events based on a combination of optimizing the exposure time,  readout-stage voltages, and fiducialization, and exploiting the elongated form factor of new detectors.

%%%%%%%%%%%%%%%%%%%%%%%%%%%%%%%%%%%%%%%%%%%%%%%%%%%%%%%%%%%%%%%%%%%%%%%%%%%%
%%%%%%%%%%%%%%%%%%%%%%%%%%%%%%%%%%%%%%%%%%%%%%%%%%%%%%%%%%%%%%%%%%%%%%%%%%%%
%%%%%%%%%%%%%%%%%%%%%%%%%%%%%%%%%%%%%%%%%%%%%%%%%%%%%%%%%%%%%%%%%%%%%%%%%%%%
%%%%%%%%%%%%%%%%%%%%%%%%%%%%%%%%%%%%%%%%%%%%%%%%%%%%%%%%%%%%%%%%%%%%%%%%%%%%
\textbf{ACKNOWLEDGMENTS.}
We thank Belina von Krosigk and Matthew Wilson for useful discussions. 
We are grateful for the support of the Heising-Simons Foundation under Grant No.~79921.
RE also acknowledges support from DoE Grant DE-SC0017938.  
This work was supported by Fermilab under DOE Contract No.\ DE-AC02-07CH11359. 
The work of TV and EE is supported by the I-CORE Program of the Planning Budgeting Committee and the Israel Science Foundation (grant No.1937/12). TV is further supported  by the European Research Council (ERC) under the EU Horizon 2020 Programme (ERC- CoG-2015 -Proposal n.~682676 LDMThExp),
and a grant from The Ambrose Monell Foundation, given by the Institute for Advanced Study.
This manuscript has been authored by Fermi Research Alliance, LLC under Contract No. DE-AC02-07CH11359 with the U.S.~Department of Energy, Office of Science, Office of High Energy Physics. The United States Government retains and the publisher, by accepting the article for publication, acknowledges that the United States Government retains a non-exclusive, paid-up, irrevocable, world-wide license to publish or reproduce the published form of this manuscript, or allow others to do so, for United States Government purposes.

 \bibliographystyle{apsrev4-1}
% \bibliography{SENSEI.bib}
 \bibliography{main.bbl}

\end{document}